# Monolithic Quantum-Dot Distributed Feedback Laser Array on Si

Yi Wang[1†], Siming Chen[2†*], Ying Yu[1], Lidan Zhou[1], Lin Liu[1], Chunchuan Yang[1], Mengya Liao[2], Mingchu Tang[2], Zizhuo Liu[2], Jiang Wu[2], Wei Li[4], Ian Ross[4], Alwyn J. Seeds[2], Huiyun Liu[2*], Siyuan Yu[1,3*]

[1]State Key Laboratory of Optoelectronic Materials and Technologies, Sun Yat-sen University, Guangzhou 510275, China.
[2]Department of Electronic and Electrical Engineering, University College London, London WC1E 7JE, UK.
[3]Department of Electronic and Electrical Eng., University of Bristol, Bristol BS8 1UB, UK.
[4]Department of Electronic and Electrical Engineering, University of Sheffield, Sheffield, S1 3JD, UK.
*Corresponding author: siming.chen@ucl.ac.uk; huiyun.liu@ucl.ac.uk; S.Yu@bristol.ac.uk

**Electrically-pumped lasers directly grown on silicon are key devices interfacing silicon microelectronics and photonics. We report here, for the first time, an electrically-pumped, room-temperature, continuous-wave (CW) and single-mode distributed feedback (DFB) laser array fabricated in InAs/GaAs quantum-dot (QD) gain material epitaxially grown on silicon. CW threshold currents as low as 12 mA and single-mode side mode suppression ratios (SMSRs) as high as 50 dB have been achieved from individual devices in the array. The laser array, compatible with state-of-the-art coarse wavelength division multiplexing (CWDM) systems, has a well-aligned channel spacing of 20±0.2 nm and exhibits a record wavelength coverage range of 100 nm, the full span of the O-band. These results indicate that, for the first time, the performance of lasers epitaxially grown on silicon is elevated to a point approaching real-world CWDM applications, demonstrating the great potential of this technology.**

## 1. INTRODUCTION

The ever-growing data volume being transported in today's on-chip and off-chip networks imposes significant challenges on copper-based interconnects. One promising approach to address this challenge is optical interconnect based on silicon photonics [1,2], which is fast maturing as a viable technology for metro and short-reach data transmission, due to the potential of low-cost, high-yield, and streamlined manufacturing enabled by the mature complementary metal–oxide–semiconductor (CMOS) fabrication technology. While the majority of photonic functions can now be realized on silicon [3], one key missing component is an efficient silicon-based laser. As an indirect-bandgap semiconductor, silicon is very inefficient in light emission [4,5]. As a result, significant efforts have been devoted to produce silicon-based lasers by integrating direct bandgap III-V compound semiconductors with silicon using either hybrid or monolithic methods. Although the former approach has been proved to be successful in producing III-V lasers and active components on silicon with device performance comparable to those grown on native III-V substrates [6-11], the monolithic approach based on epitaxial growth, in the longer term, is more desirable when it comes to mass production, for its potential in realizing low-cost, high-yield, and reliable manufacturing processes.

However, monolithic growth of III-Vs on Si faces considerable hurdles including large lattice mismatch (4% for GaAs and 8% for InP), difference in thermal expansion coefficients, and polar versus non-polar surfaces, which result in several types of defects such as threading dislocations (TDs), micro-thermal cracks and antiphase boundaries (APBs), respectively, severely degrading the promise of III-V materials [12]. III-V quantum-dot (QD) structures have become one of the most promising material systems for III-V semiconductor lasers to achieve ultimately superior device performance [13-15]. Coincidentally, QDs have also been proved to be less sensitive to defects than conventional quantum well (QW) structures due to effective carrier localization [16]. It is, therefore, no surprise that researchers in the field of silicon photonics have strived for years to develop monolithically integrated III-V QD light sources on silicon substrates that can exploit the benefits of QD while also being able to enjoy fully the economics of scale promised by monolithic growth. As a result, III-V QD Fabry-Perot (FP) lasers monolithically grown on Si have made significant progress and have outperformed their QW and bulk counterparts in terms of lower threshold current, higher temperature insensitivity, higher efficiency and longer lifetime [16-25].

While single channel optical interconnects can operate with silicon-based electrically-pumped FP laser sources, high-performance applications would require arrays of multiple single-mode lasers with carefully designed cavities that enable wavelength division multiplexing (WDM) [26]. Recently, leveraging a novel selective-area growth technique in confined V-shaped grooves, the growth of high-quality InP-based material directly on buffer-free silicon has been achieved [27]. While the first room-temperature monolithic Si-based InP distributed feedback (DFB) laser array emitting within different wavelength ranges has been reported using this method [28,29], the array can only operate by optical pumping and its wavelength coverage is limited to 40 nm [29]. In this work, we take another step forward and demonstrate, to the best of knowledge, the first electrically-pumped DFB laser array using InAs/GaAs QD gain material epitaxially grown on silicon. The array operates under continuous-wave (CW) conditions at room temperature, and features threshold currents as low as 12 mA, side mode suppression ratios (SMSRs) as high as 50 dB and a record wavelength covering range of 100 nm the full span of the O-band, demonstrating the great potential of the monolithic approach in real systems applications.

## 2. DESIGN AND FABRICATION

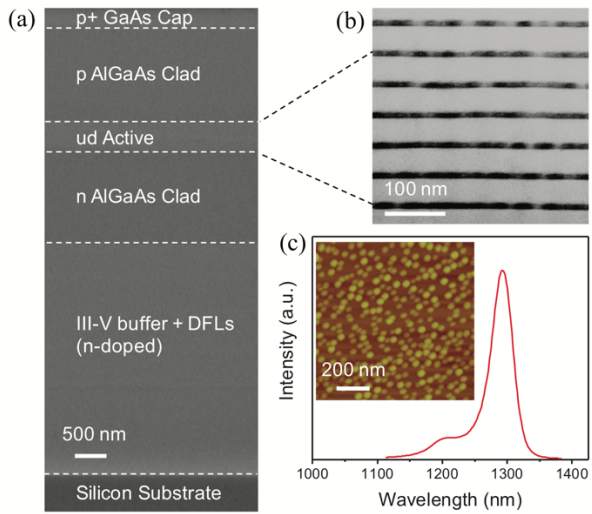

**Fig. 1.** The material properties of Si-based QD lasers. (a) SEM image of the transverse layer structure of the epi-wafer used in laser fabrication. (b) Bright field STEM image of the active layers. (c) Photoluminescence spectrum of the QD active layers on silicon peaking at 1297nm. The inset shows the atomic force microscope image of an uncapped QD layer.

In this work, an InAs/GaAs QD laser structure [Fig. 1(a)] was directly grown on an n-doped silicon (001) substrate with a 4° off-cut angle towards the [011] plane. To achieve high-quality lasers epitaxially grown on silicon, it is crucial to minimize the impact of TDs generated due to the large lattice mismatch between III-Vs and silicon [5, 30-33]. Here, special growth techniques have been developed, in which an AlAs nucleation layer, InGaAs/GaAs dislocation filter layers (DFLs) [19] and *in-situ* thermal annealing [34] have been utilized following previous optimized conditions [21], to deliver high-quality III-V buffer layers grown directly on Si with low TD density of around $1.2 \times 10^6$ cm$^{-2}$ determined by transmission electron microscopy (TEM). Following the III-V buffer layers, a standard p-i-n laser structure was grown. The active region was composed of a 7-layer InAs/InGaAs dots-in-a-well (DWELL) structure separated by 50 nm GaAs spacer layers. Each DWELL layer consists of 3 monolayers of InAs sandwiched between 2 nm lower and 6 nm upper In$_{0.15}$Ga$_{0.85}$As layers [21]. By virtue of the low TD density in the III-V buffer layers, a near defect-free active region [Fig. 1(b)] was achieved. A room-temperature photoluminescence (PL) emission peaking at 1297 nm was observed with a narrow linewidth of 30 meV, as seen in Fig. 1(c). A good QD uniformity and a QD density of $\sim 3 \times 10^{10}$ cm$^{-2}$ were achieved, as indicated by the atomic force micrograph from an uncapped QD test sample grown on silicon under the same growth conditions. The 3-inch wafer was then diced into smaller pieces and fabricated into broad-area FP lasers (Supplementary Section 1) and the DFB laser array described in this paper.

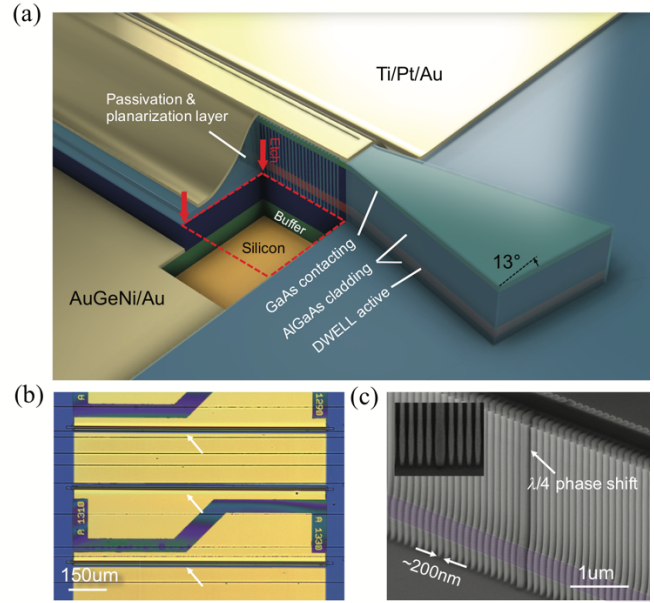

**Fig. 2.** The DFB laser array on silicon. (a) Cutaway schematic showing the vertical layer structure, the output coupler and the etched gratings (not to scale). (b) Regional microscope image of the DFB laser array on silicon. The arrows indicate positions of the ridge waveguides. (c) High-resolution SEM image of the gratings with a $\lambda/4$ phase shift in the middle from a previous test run. The e-beam resist is still present. Shaded in light purple is the active region. Inset: the SEM image of the gratings from a near 90-degree viewpoint showing the high-quality gratings with almost no residue. The scale bar applies to both images.

Conventionally, development of a III-V DFB laser normally involves two epitaxial growth sequences, one for the lower cladding and the active region, the other for the upper cladding and the contact layer following the fabrication of shallow-etched gratings near the waveguide core [35]. Alternatively, a simplified method avoiding complicated regrowth has been developed, in which lateral surface gratings are fabricated simultaneously in the waveguide etching process [36,37]. The gratings are etched to just above the active region to achieve strong photon-grating interactions as well as low optical loss. For InP-based structures, an aluminum-containing stop-etch layer and a chemically selective etching recipe are often used together to ensure the etch depth to be precisely above the active region, so as to obtain an accurate grating coupling coefficient $\kappa$. However, in the InAs/GaAs QD material system described above, such a chemically selective etch-stop layer has not been found to date. Therefore, in our design, waveguides with lateral gratings engraved on their sidewalls that penetrated the active region were used. To retain the high repeatability, these were intentionally over-etched to about 500 nm into the lower cladding, as seen in Fig. 2(a). In doing so, an inductively coupled plasma (ICP) etching recipe of ultra-high aspect ratio (>30:1) using the SiCl$_4$/N$_2$ chemistry was developed. Fig. 2(c) shows the high-resolution scanning electron microscope (SEM) image of 3.4 μm deep etched gratings with masks generated by e-beam lithography (EBL). Exploiting such deep-etching method could enhance the immunity to the variations of the etch depth expected in different etching batches, which would otherwise result in drift of the effective refractive index leading to detrimental wavelength shifting (Supplementary Section 2). In addition, as QDs were used in the active region, the negative impact of non-radiative recombination occurring on the ridge surface has been minimized, owing to its improved insensitivity to defects. A simple yet robust fabrication process was therefore achieved.

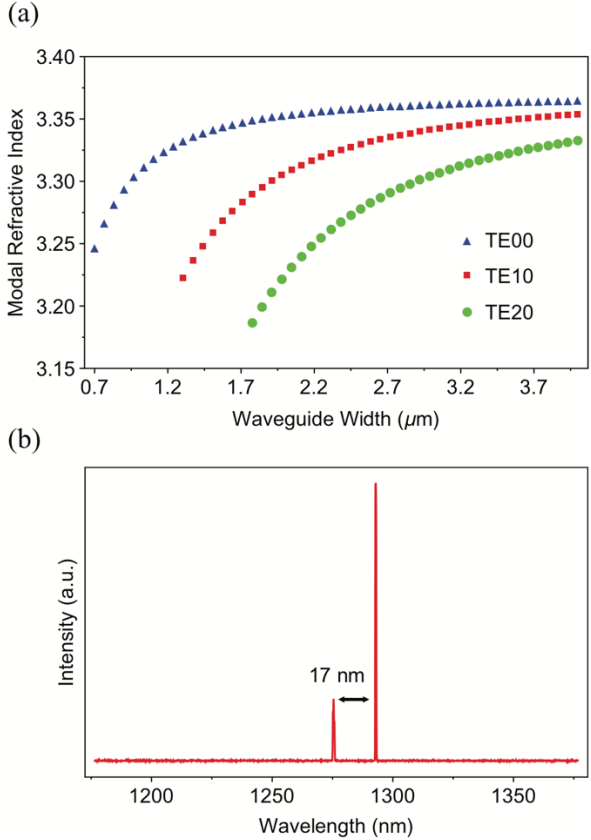

**Fig. 3.** Modal analysis of the DFB laser. (a) Modal refractive indices of the first three modes in deep-etched waveguides of different widths. (b) Optical spectrum of a multimode DFB laser with a waveguide width of 2.2 μm and total length of 1.5 mm, operating above threshold with continuous electrical pumping at room temperature.

Typically, sufficiently narrow waveguides are used in DFB lasers to support only the fundamental transverse mode in order to guarantee single-transverse-mode operation. However, given the deep-etched waveguide design, the single-mode waveguide is too narrow to provide sufficient gain, or to ensure a reasonable grating size for good reproducibility. In our design, a waveguide width of 2.2 μm supporting the first three transverse modes, as seen in Fig. 3(a), was found to be a good trade-off by showing single-mode lasing in the 2nd-order mode. Having a larger grating coupling coefficient $\kappa$ compared to the fundamental mode and, being still well confined in the waveguide thanks to the sufficiently wide ridge, the 2nd-order transverse mode can reach threshold first and suppress the fundamental mode from lasing thereafter. Theoretically, $\kappa$ can be estimated by the difference of refractive indices between two transverse slices of the waveguide [38,39], one slice on the peak and the other on the trough. Furthermore, in contrast to the 2nd-order mode, the 3rd-order mode is more sensitive to surface scattering due to poor confinement which eventually prevents it from lasing. No signs of lasing in the 3rd-order mode were found experimentally.

The experimental evidence of lasing in the 2nd-order mode was found using DFB lasers with a dimension of 2.2 μm × 1.5 mm. By electrically pumping these lasers to above threshold, 2-mode lasing with an extra mode sitting at a wavelength 17 nm longer than the desired mode was observed [Fig. 3(b)], which was in good agreement with the theoretical spacing of 15.5 nm between the $TE_{10}$ and $TE_{00}$ mode. By comparison, the theoretical wavelength spacing between $TE_{20}$ and $TE_{10}$ mode was 26.3 nm. We therefore deduced that the two lasing peaks corresponded to the $TE_{10}$ and $TE_{00}$ mode, respectively. To eliminate the unstable $TE_{00}$ mode, the total length of the grating was decreased to 1.2 mm and 1 mm, thus lowering the total coupling strength and increasing the threshold for the fundamental mode. This design provides robust single mode operations, which will be shown later in the paper. It should be noted that the 2nd-order mode can be easily converted into the fundamental mode by waveguide mode converters [40], for the purpose of coupling into single mode waveguides in future versions of the device.

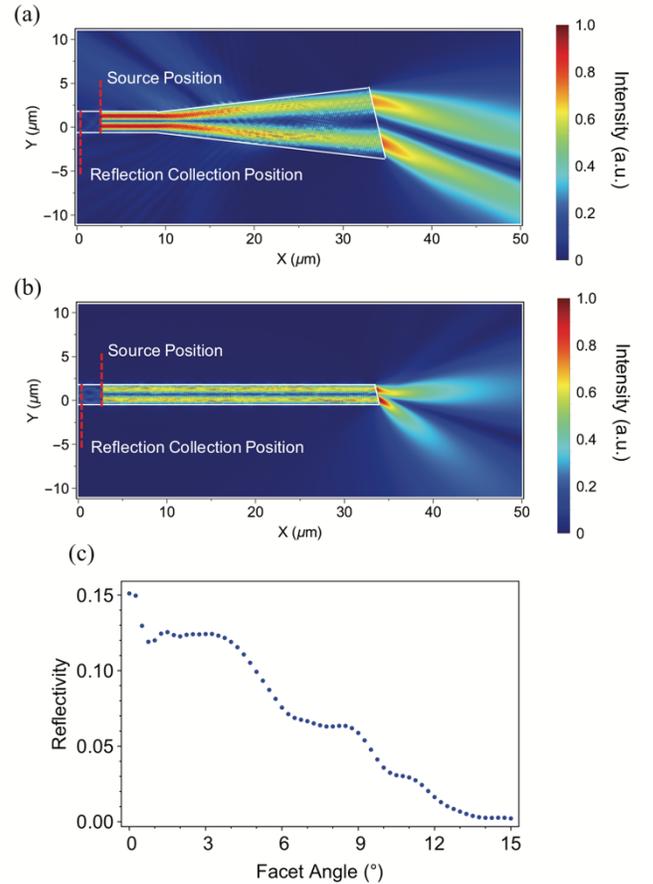

**Fig. 4.** FDTD simulation result of the output coupler with a $TE_{10}$ modal input showing low reflection back into the waveguide. The waveguide outline, the source position and the reflection collection position are drawn in the figure respectively. (a) The angled output facet with a forward taper. The reflectivity is ~1%. (b) The angled output facet without a forward taper. The reflectivity is ~10%. (c) FDTD simulated reflectivity of the AR output coupler with an 8 μm wide, 25 μm long forward taper attached to the 2.2 μm deep-etched waveguide.

To achieve a suitable $\kappa$, low radiation loss, and a reliable fabrication process, the waveguide-embedded sidewall gratings have been designed as 50:50 first-order, with an intrusion of 100 nm. A $\lambda/4$ phase shift was placed in the middle of the gratings to force single-longitudinal-mode lasing in the defect mode. The WDM-compatible laser array was achieved by carefully adjusting the grating periods of the lasers in the array. The 1st-order grating period values $p$ were determined first coarsely, using $n_{eff}$ obtained by numerical simulations and the relation

$$p = \frac{\lambda}{2n_{eff}}. \quad (1)$$

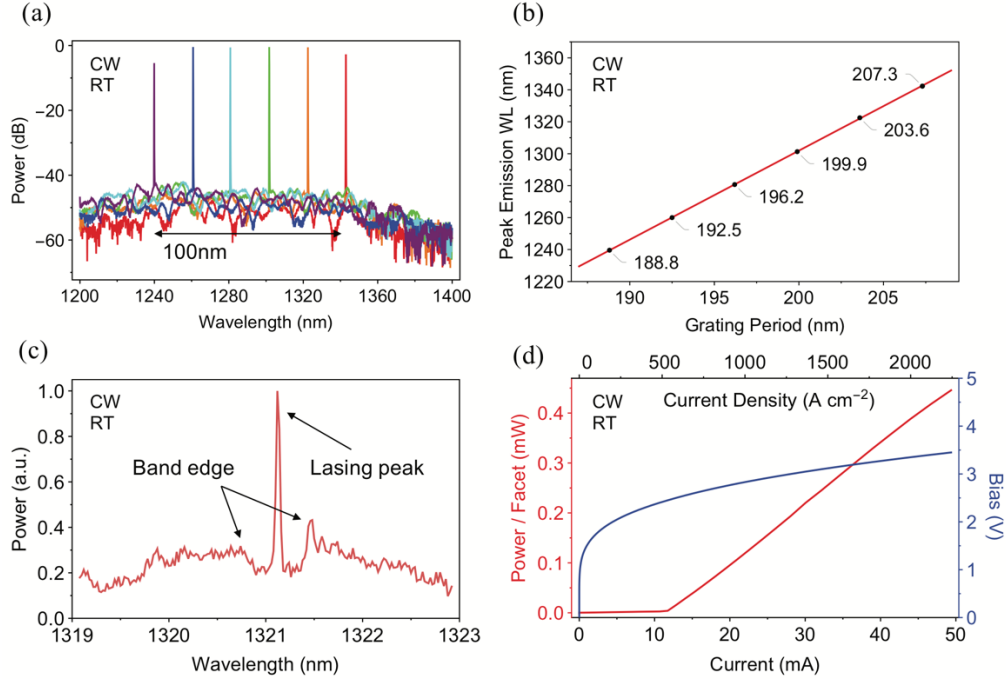

**Fig. 5.** Continuous-wave test results of the silicon-based DFB laser array at room temperature. (a) Optical spectra of a DFB laser array with different grating periods around their maximum output power levels before saturation at room temperature. Resolution: 0.1 nm. (b) Peak emission wavelengths of the DFB laser array plotted against grating period values. The labels indicate the corresponding grating period values of the lasers in the array. The wavelengths were obtained at the same current density of 1.9 kA cm$^{-2}$. (c) Zoomed-in optical spectrum of a particular DFB laser operating just below threshold. Resolution: 0.07 nm. (d) Light-current-voltage curve of a single 1 mm-long silicon-based DFB laser. The output power was collected at one of the two symmetric facets.

As seen in the equation, any variations of $n_{eff}$ will lead to corresponding changes in the lasing wavelength $\lambda$ assuming fixed grating periods, which is inevitable with the presence of simulation and fabrication errors. However, given the linear relation between $\lambda$ and $p$, more accurate grating period values can be determined from a calibration batch on the same wafer, including fabrication of coarsely-designed DFB lasers and measurement of their lasing wavelengths.

Another important factor for DFB lasers is the suppression of facet back reflections, and failure to do so may cause mode jumping at high currents. Here, owing to the high-aspect-ratio etching process, this problem was tackled simply and cost-effectively by dry-etched anti-reflection (AR) output couplers instead of additional AR coatings. Generally, by tilting the output facet and widening the waveguide, the facet reflectivity can be greatly reduced due to mode mismatch in the reflected light [41]. However, large tilting angles and waveguide widths may cause difficulties in coupling. As a trade-off, the combined use of an 8 μm wide, 25 μm long forward taper and a 13°-tilted facet successfully decreased the reflection coefficient down to ~1% and increased the output efficiency up to ~82%, meanwhile maintaining a reasonable beam profile that can be collected by an optical fiber. The simulation results were obtained by finite-difference time-domain (FDTD) simulations, as shown in Fig. 4. The absorption of the un-pumped output coupler was estimated to be 7% by FDTD simulations, and this was considered acceptable. Successful fabrication of the facet was supported by the high SMSR and high mode stability demonstrated later in the paper. Note that the use of etched facets can also avoid the issue of unpredictable cleaving of off-cut silicon substrates and, more significantly, pave the way towards monolithic laser-waveguide integration by enabling high-performance devices without cleaving. A detailed description of the fabrication process can be found in the Methods section of supplementary materials.

## 3. RESULTS

To analyze the optical spectrum of the silicon-based WDM DFB laser array, the finished wafer was first diced into independent bars each containing multiple DFB lasers with varying grating periods, and then placed face-up on a 3-axis aligning stage for probe-testing. The individual lasers on the bar were biased with a direct current (DC) source at room temperature (24°C) with no active cooling, and then butt-coupled with a 50 μm/125 μm multimode fiber. CW single-mode lasing was observed. The 6-device array produced a wavelength range of 100 nm around the 1300 nm communication band, as seen in Fig. 5(a). A 0.1 nm precision was achieved for the grating period, yielding a channel spacing of 20±0.2 nm, matching well with the standard coarse wavelength division multiplex (CWDM) grid.

A typical λ/4 phase-shifted DFB laser spectrum is shown for sub-threshold operation on a particular laser device in Fig. 5(c). The spectrum consists of a 0.8 nm-wide bandgap and a central defect mode which would lase at higher currents. Assuming infinitely long gratings, the total coupling strength $\kappa L$ for the 1.2 mm-long device was estimated by the bandgap width to be around 5. This relatively strong coupling strength could alleviate any residual facet back reflections introduced by fabrication errors, thus improving the single mode quality. The mode stability is also evident in the light-current-voltage (LIV) curve in Fig. 5(d), where above the 550 A cm$^{-2}$ (12 mA) threshold the output power follows a kink-free near-linear curve for the 1 mm-long silicon-based DFB laser. The slope efficiency and wall-plug efficiency at 49 mA was calculated to be 0.024 W/A and 0.5% respectively. As the p- and n-electrodes were both fabricated on the epi-side, the current flow could avoid the defect-rich III-V/Si interface, and therefore a slope resistance of around 20 Ω after diode turn-on was achieved. Room-temperature output power per facet exceeded 0.5 mW and 1.5 mW under CW and pulsed conditions of 1 μs pulse width and 1% duty ratio, respectively. More information regarding the LIV curves of the array can be found in Supplementary Section 3.

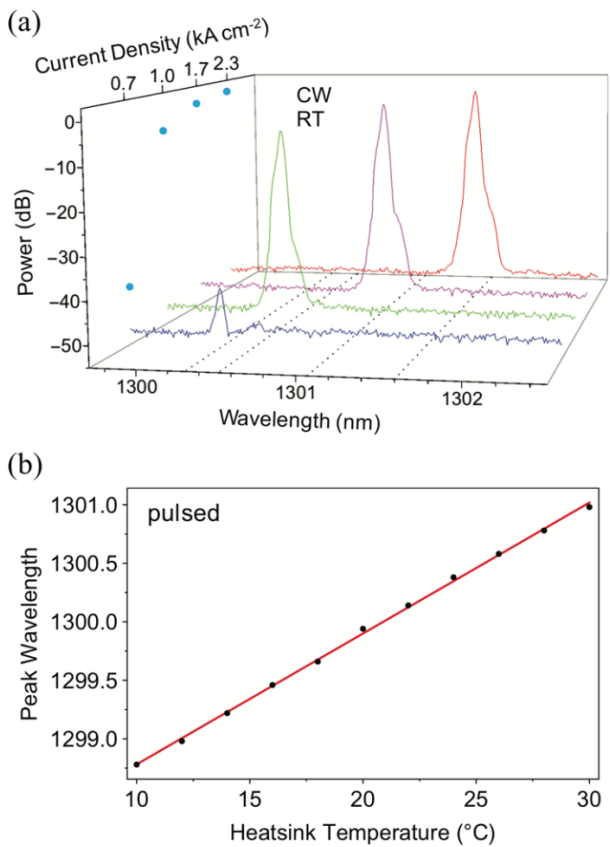

**Fig. 6.** Mode stability test results on one silicon-based DFB laser in the array. (a) Normalized peak power (left panel) and optical spectra of a single DFB laser with different DC currents at room temperature. (b) Linear fit of peak emission wavelengths for the DFB laser with a pulsed source of 1 μs pulse width and 1% duty cycle at different heatsink temperatures.

The mode stability of communications application silicon-based DFB lasers has also been characterized in terms of SMSR and temperature-dependent or current-dependent wavelength shift. As plotted in Fig. 6(a), starting from a weak amplified spontaneous emission (ASE) below threshold, the lasing mode later reached a high SMSR of ~50 dB at 2.3 kA cm$^{-2}$ (60 mA). The high SMSR value implies a correct combination of the design parameters and the successful fabrication of etched AR facet. By varying the injection currents, the peak emission wavelength drifted at a rate of 0.03 nm/mA or 8.6 pm/mW corresponding to a shift of 1.2 nm at 2.3 kA cm$^{-2}$ (60 mA), which was well within the CWDM channel window. Temperature-dependent measurement was carried out under a pulsed condition of 1 μs pulse width and 1% duty ratio, to minimize the effect of electrical self-heating. As shown in Fig. 6(b), the wavelength thermal drift was fitted to be 0.11 nm/°C, which meant the CW device temperature at 60 mA was around 35 °C (11°C above substrate temperature). The thermal resistance calculated was approximately 50 °C/W. It is worth mentioning that the data presented above represent the worst-case results, because the lasers were operated epi-side up without substrate thinning (the substrate thickness was 410 μm) and they were not hard-soldered to a high thermal conductivity heatsink; The lasers were also directly probed without wire-bonding. Proper bonding would improve the device performance. The high performance obtained even in such harsh testing conditions further confirmed the potential of this laser array in real application scenarios with demanding environmental requirements.

## 4. CONCLUSION

We have designed and demonstrated, for the first time to the best of our knowledge, an electrically-pumped CWDM DFB laser array using InAs/GaAs QD gain materials monolithically grown on a silicon substrate. Operating at room temperature, the DFB laser array exhibited threshold currents as low as 12 mA, SMSRs as high as 50 dB and a wavelength coverage of 100 nm with a precise channel spacing of 20±0.2 nm. These results meet the CWDM requirements and represent a major step towards fully monolithic silicon-based photonic integrated circuits (PICs) for low-cost and high-performance applications. The Si-based DFB laser technology can also be used for non-communications applications such as on-chip sensing and metrology where integrated single mode coherent sources are a key part of the functional PICs.

**Funding**. UK Engineering and Physical Sciences Research Council (EPSRC) (EP/J012904/1, EP/J012815/1.); National Natural Science Foundation of China (NSFC) (61490715).

**Acknowledgement.** S.C. thanks the Royal Academy of Engineering for funding his Research Fellowship under Ref No. RF201617/16/28.

See Supplement 1 for supporting content.

†These authors contributed equally to this work.